# KEK EFFORT FOR HIGH FIELD MAGNETS

T. Nakamoto, KEK, Tsukuba, Japan.

*Abstract*

KEK has emphasized efforts to develop the RHQ-$Nb_3Al$ superconductor and a sub-scale magnet reaching 13 T towards the HL-LHC upgrade in last years. In addition, relevant R&D regarding radiation resistance has been carried out. For higher field magnets beyond 15 T, HTS in combination with A15 superconductors should be one of baseline materials. However, all these superconductors are very sensitive to stress and strain and thorough understanding of behaviour is truly desired for realization of high field magnets. KEK has launched a new research subject on stress/strain sensitivity of HTS and A15 superconductors in collaboration with the neutron diffraction facility at J-PARC and High Field Laboratory in Tohoku University. Present activity for high field magnets at KEK is reported.

## INTRODUCTION

Recently, future upgrade of the LHC has been lively discussed. The high luminosity LHC upgrade (HL-LHC), which would provide 5 times higher beam luminosity than the present LHC, has been discussed as the most possible upgrade plan in near future. Present beam insertion systems for ATLAS and CMS will be replaced by new superconducting magnets to attain smaller β* with a larger beam aperture. A15 type superconductors such as $Nb_3Sn$ and $Nb_3Al$ can generate higher field up to 15 T and are considered to be promising materials for the HL-LHC.

Beyond the HL-LHC, the high energy LHC (HE-LHC) might be realized in 2030 or later. In the HE-LHC, the beam energy is expected to be double at least and nominal field of new main dipole magnets in the LHC tunnel should reach 20 T or more. This means that utilization of HTS (high temperature superconductors) in combination with A15 type superconductors needs to be considered. However, it is very well known that these superconductors are brittle and superconducting performance such as critical current density $J_c$ is significantly influenced by mechanical stress and strain. Comprehension of these effects is definitely necessary to realize high field superconducting magnet for future accelerator.

KEK has been engaged to develop the $Nb_3Al$ superconductors and the high field magnet technology towards the future accelerator. Recently, we have also launched new research subject regarding the stress and strain sensitivity of the superconductors. In this paper, present R&D status and future plan at KEK towards the high field magnet are reported.

## PRESENT R&D STATUS

Under the framework of the CERN-KEK collaboration, KEK has developed the $Nb_3Al$ superconductor for the high field accelerator magnet application. This R&D work is complementary to other R&D efforts in CERN and US-LARP with the $Nb_3Sn$ superconductors.

A tentative target application is set to the HL-LHC where the magnets below 15 T would be utilized. In parallel with the superconductor development, KEK has been developing a sub-scale magnet to demonstrate feasibility of $Nb_3Al$ cable. KEK has been also performing R&D on the relevant magnet technologies such as insulations, radiation resistance study.

### Nb3Al Superconductor Development

For the accelerator magnet application beyond 10 T, $Nb_3Sn$ superconductor is in the most advanced state of development. However, $J_c$ can be degraded by excessive stress and strain. In contrary, $Nb_3Al$ has a much better stress and strain tolerance. For instance, the previous study demonstrated that an $Nb_3Al$ strand in an epoxy-impregnated cable sustained under transversal stress beyond 200 MPa [1].

In order to utilize the better strain tolerance, studies on the development of $Nb_3Al$ wires have been conducted in Japan for many years. Thanks to a Rapid Heating/Quenching and transformation (RHQ) process [2] developed by NIMS (National Institute for Materials Science), $J_c$ has been significantly improved at high field region. However, the wire temperature instantaneously reaches around 2000 °C to form supersaturated solid solution of Nb(Al)ss in the RHQ process and an ordinary copper matrix cannot be utilized because it melts. Therefore, main development items of RHQ-$Nb_3Al$ wires for accelerator application are not only to increase of $J_c$ but also adoption of adequate matrix and stabilizer.

KEK and NIMS have been jointly developing RHQ-$Nb_3Al$ wires for the accelerator application. Figure 1 shows a cross section of the recent RHQ-$Nb_3Al$ wire and the specification is listed in Table 1.

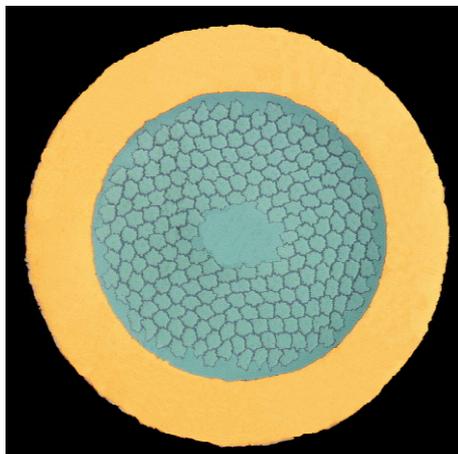

Figure 1: Cross section of the RHQ-$Nb_3Al$ superconducting wire

Table 1: Specification of the RHQ-Nb$_3$Al Wires

| | |
|---|---|
| Wire Diameter | 1.0 mm |
| Non-Copper Diameter | 0.7 – 0.73 mm |
| Area Reduction | ~70 % |
| Filament Diameter | 35 μm |
| Barrier Thickness | 4 - 6 μm |
| Twist Pitch | 45 mm |
| Piece Length | ~ 1 km |

Firstly, an ordinary Nb-matrix that has better mechanical affinity to other composite elements was incorporated. The RHQ-Nb$_3$Al wire with Nb-matrix showed the highest non-copper $J_c$ of 1030 A/mm$^2$ at 15 T. Furthermore, wire breaking rate during cold drawing was rather small. However, it turned out that Nb-matrix wires exhibited large magnetic instability in a low-field region at 4.2 K where niobium is in the superconducting state. It was concluded that utilization of Nb-matrix was not appropriate for the accelerator application.

Accordingly, new Ta-matrix wires that are stable in a low field have been developed. Figure 2 shows magnetization curves of RHQ-Nb$_3$Al wires with Nb- and Ta-matrix. In comparison with Nb-matrix wire (F1), magnetization curves of Ta-matrix wires (K1 and K2) are very small and no flux jump can be seen. In terms of suppression of low field instability, adoption on Ta-matrix was very successful. Figure 3 shows non-copper Jc of Ta-matrix wires (K1-K4). Although each wire has design parameters in the cross section, behaviours of non-copper Jc are very similar. However, an average non-copper Jc at 15 T is still around 800 A/mm$^2$ and this value is about half of Nb$_3$Sn (OST-RRP).

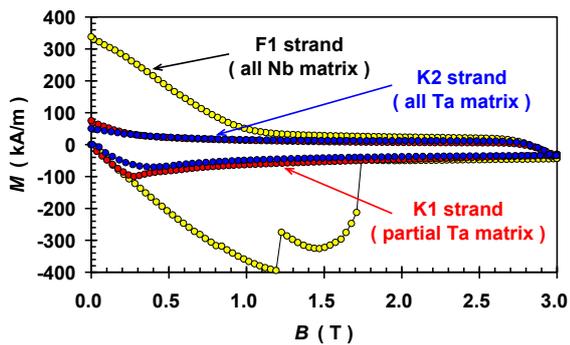

Figure 2: Magnetization curves of RHQ-Nb$_3$Al wires at 4.2 K. [3]

Regarding manufacturing of the Ta-matrix wires, we have suffered wire breakings during cold drawing so far because tantalum is rather stiff. Microscopic observation indicated that the breaking was initiated at very narrow tantalum matrix in the cross section. In order to reduce the breaking rate towards 10 km class long wire production, importance of quality improvement and control for tantalum sheets was recognized. Production trials of 100 m long wire using different tantalum ingredients were started in 2010. The study is still underway and the successful results would be adopted for the 1 km long wire production as a demonstration in 2011.

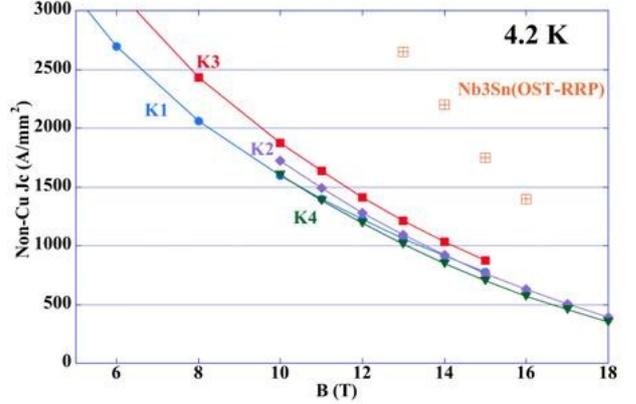

Figure 3: Non-copper $J_c$ of RHQ-Nb$_3$Al wires with Ta-matrix. As a reference, non-copper $J_c$ of Nb$_3$Sn-RRP is also plotted.

*Nb$_3$Al Sub-Scale Magnet Development*

In parallel with the superconductor development, KEK has made progress on the development of a 300 mm long Nb$_3$Al superconducting sub-scale magnet with a simple mechanical structure that is considered to be a fundamental R&D to demonstrate feasibility of high field magnets with Nb$_3$Al cable [4]. So far, four types of Nb$_3$Al Rutherford cables for the sub-scale magnet have been successfully fabricated in collaboration with Fermilab. The cable with 28 RHQ-Nb$_3$Al strands is 14 mm wide and 1.8 mm thick and a piece length is over 20 m.

Figure 4 shows a schematic drawing of the sub-scale magnet. The design concept is incorporated from the original development by LBNL with Nb$_3$Sn technology. The magnet consists of three Nb$_3$Al coils combined with two Nb$_3$Sn coils [5]. The magnet has the "common-coil" structure with a very narrow gap along the vertical median plane such that the peak field in the central Nb$_3$Al coil can be maximized to reach 13.2 T at 12 kA. Two additional Nb$_3$Sn sub-scale coils developed by LBNL with a higher current density contribute to boost up the peak field effectively. A rather thick aluminium shell is required to apply adequate pre-stress at magnet assembly and its large thermal shrinkage can increase the pre-stress even during cool-down.

Following two dummy coil fabrication with NbTi cables to evaluate the fabrication process including heat reaction in the vacuum furnace at 800 °C and the epoxy resin vacuum impregnation, the first real coil winding with Nb$_3$Al cable was carried out. Another two coils will be fabricated in 2011.

*Relevant Magnet Technology R&D*

In superconducting magnets for the HL-LHC with Nb$_3$Al superconductor, the coil insulation system plays very important role. The insulation system needs to fulfil the following specification: endurance at higher reaction temperature than $Nb_3Sn$, mechanical reinforcement by resin impregnation under very severe radiation environment, and keeping engineering current density as high as possible.

Since heat reaction temperature of 800 °C for the $Nb_3Al$ coil is higher than $Nb_3Sn$, an ordinary glass tape in the $Nb_3Sn$ system are not applicable. Two types of alumina tape of 0.125 mm thick from different suppliers (CTD and NITIVY) have been used for the cable insulation. Recently, NITIVY has succeeded to manufacture thinner alumina tape of 0.08 mm thick aiming higher engineering current density. This insulation will be utilized for the coil winding in near future.

In terms of radiation resistance, an ordinary epoxy resin that is commonly used for the $Nb_3Sn$ coil impregnation is only applicable up to several MGy. Cyanate ester is known to have much better radiation resistance than epoxy. However, higher curing temperature beyond 180 °C and extension of pot life with low viscosity are practical issues for the coil impregnation. KEK and other three institutes (JAEA, University of Hyogo and Mitsubushi Gas Chemical) have formed a collaboration framework and have newly developed the special cyanate ester base resin mixed with epoxy for the accelerator coil application: lower curing temperature at 150 °C and long pot life of 24 hours at 60 °C. Demonstration of the dummy coil impregnation with the cyanate ester base resin was successfully carried out and the picture of the impregnated coil is shown in Fig. 5.

Evaluation of radiation resistance of magnet materials is crucial. We have carried out gamma ray irradiation tests on organic materials since 2003 at JAEA-Takasaki. The new cyanate ester resin is planned to be evaluated soon. In addition, a series of neutron irradiation tests at cryogenic temperature below 20 K have been launched in 2010 at KUR (Kyoto University Research Reactor). Main scope is to survey electric resistivity increase of stabilizers due to neutron irradiation: resistivity of pure metal is known to be degraded even at fast neutron fluence of $10^{21}$ $n/m^2$ or less, but the data for industrial stabilizers such as aluminium and copper with RRR up to several 100 does not exist. The quench protection scheme of the magnet system is concerned to be compromised when the resistivity of the stabilizer unexpectedly increases due to the neutron irradiation during the beam operation. The first irradiation test with aluminium samples from the superconducting cable for solenoid magnets have been made and rapid increase of resistivity has been observed even at neutron fluence of $10^{20}$ $n/m^2$. The next irradiation test for copper stabilizers for the accelerator magnets is planned in 2011.

## NEW RESEARCH ON STRESS/STRAIN DEPENDENCE OF SUPERCONDCUTORS

For high field accelerator magnets beyond 15 T in future, utilization of HTS and A15 type superconductors is considered as baseline materials in the meantime. As mentioned above, however, it should be reminded that their performance like $J_c$ depends on stress and strain of superconductors. The engineering design for such high field magnets must need thorough understanding of stress/strain dependence of the superconductor performance.

Since industrial superconducting wires/tapes are composites comprised of superconductors surrounded by other materials and stabilizers with different thermal contractions, residual strains can be naturally induced by a temperature variation of around 1000 K from the heat reaction temperature to the cold for the operation. In addition, since the shape of the Rutherford-type cable is complicated and the impregnated coil windings for the accelerator magnets are applied the complicated stress in various directions during the assembly, the cool-down and the excitation, it is very difficult to predict the actual strain of the superconductor. To design and develop the high field superconducting accelerator magnets successfully, it is very important to understand the strain behaviors of the superconductor by neutron diffraction measurements. The neutron diffraction measurement facility using pulsed neutrons at the BL-19 (TAKUMI) of J-PARC MLF, shown in Fig. 6, is the most appropriate tool to experimentally study the strain behaviors of the HTS and the A15 superconductor in the accelerator coil. The following is the main reasons;

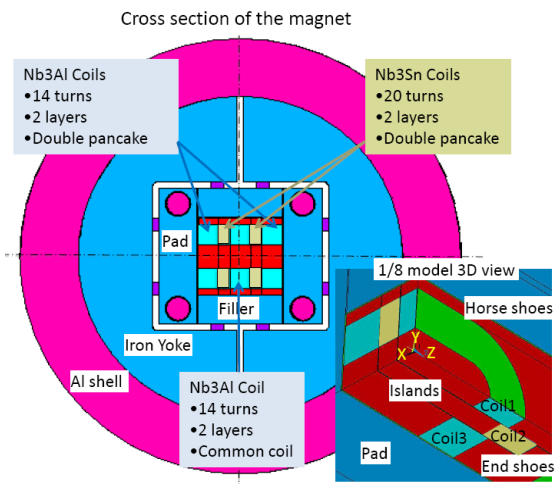

Figure 4: A schematic drawing of the $Nb_3Al$ and $Nb_3Sn$ hybrid sub-scale magnet.

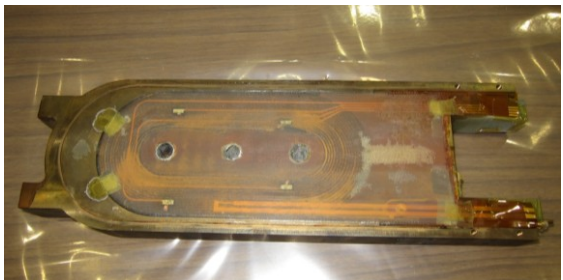

Figure 5: Picture of a dummy sub-scale coil impregnated with the new cyanate ester base resin.

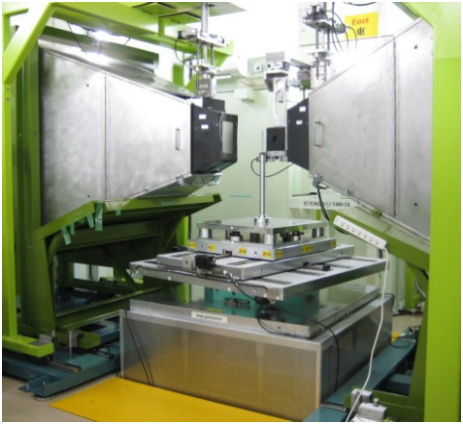

Figure 6: Neutron diffraction measurement facility at the BL-19 (TAKUMI) of J-PARC MLF.

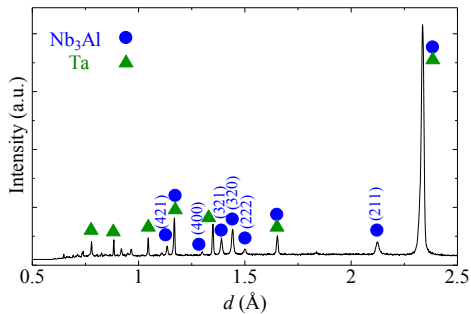

Figure 7: Diffraction peaks of RHQ-$Nb_3Al$ with Ta-matrix measured at TAKUMI, J-PARC MLF.

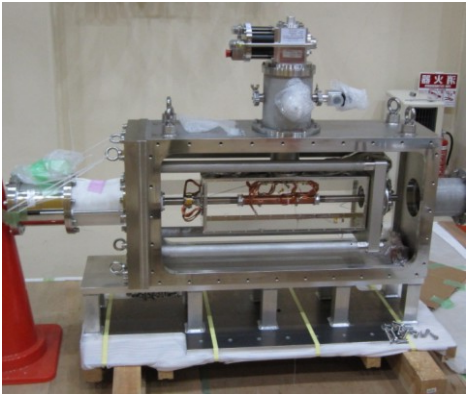

Figure 8: Cryogenic load frame up to 50 kN from 6 K to 300 K.

- Three-dimensional strains of the superconducting wire can be simultaneously determined by using a pair of 90° detector banks.
- Strains of each ingredient can be determined by using several diffraction peaks measured simultaneously.
- Even small strains can be measured with its high resolution.
- Thanks to high penetration depth of neutrons into the sample, strain distribution inside the massive coil sample can be obtained.

In collaboration with NIMS, JAEA and Tohoku University, a series of neutron diffraction measurement for the stress/strain study has been started with a long-term viewpoint. Preliminary measurement of the RHQ-$Nb_3Al$ wires with different matrixes at room temperature was made in 2010. Figure 7 shows diffraction peaks of the Ta-matrix wire as a typical case. It was observed that residual strains of $Nb_3Al$ crystal were varied according to matrix materials.

For the neutron diffraction measurement at cryogenic temperature under loading, KEK in collaboration with JAEA has newly developed a cryogenic load frame that can apply the tensile load up to 50 kN in the temperature range from 6 K to 300 K, shown in Fig. 8. This cryogenic load frame can provide different conditions to the samples with changing the load and the temperature. Not only sole superconducting wires or tapes, but also bulk samples like epoxy-impregnated Rutherford cable stacks simulating the coil can be measured with this cryogenic load frame.

In parallel, experimental study on $J_c$ behaviors under different stress/strain has been also started at High Field Laboratory in Tohoku University. Both experimental results of neutron diffraction measurement and Jc measurement under stress/strain will be inseparably analyzed. Knowledge and understanding from this study will improve the mechanical design of high field superconducting accelerator magnets and help to precisely predict its performance limit.

## SUMMARY

KEK has promoted the R&D towards high field accelerator magnet. Development of RHQ-$Nb_3Al$ superconductors aiming to be applied for the HL-LHC has been emphasized. The latest Ta-matrix wire showed better low field instability even though non-copper Jc is smaller than that of $Nb_3Sn$-RRP. Magnet technology for RHQ-$Nb_3Al$ cable under very severe radiation environment is underway. Especially, development of the cyanate ester resin for the accelerator application is highlighted. For the long-term R&D, experimental study on stress/strain sensitivity has been launched.